\documentclass[sigconf]{acmart}
\AtBeginDocument{%
	\providecommand\BibTeX{{%
			\normalfont B\kern-0.5em{\scshape i\kern-0.25em b}\kern-0.8em\TeX}}}
		
\acmConference[KDD 2023]{}{August 6, 2023}{Long Beach, California, United States}

\usepackage{ mathdots  }
\usepackage{ amsmath   }
\usepackage{ amssymb   }
\usepackage{ color     }
\usepackage{ graphicx  }
\usepackage{ amsfonts  }
\usepackage{ epstopdf  }
\usepackage{ upgreek   }
\usepackage{ bm        }
\usepackage{ graphicx  }
\usepackage{ subfigure }
\usepackage{ mathtools }
\usepackage{ algorithm }
\usepackage{algorithmic}
\usepackage{ booktabs  }
\usepackage{ multirow  }
\usepackage{ caption   }
\usepackage{ array     }
\usepackage{ float     }
\usepackage{   soul    }
\usepackage[normalem]{ulem}
\usepackage{tablefootnote}

\usepackage{bbm}
\usepackage[normalem]{ulem}
\useunder{\uline}{\ul}{}

\newtheorem{Theorem}{Theorem}

\newtheorem{Assumption}{Assumption}[section]
\newtheorem{Remark}{Remark}
\newtheorem{Proposition}{Proposition}

\newcommand{\bx}{\boldsymbol{x}}

\newcommand{\bmu}{\boldsymbol{\mu}}
\newcommand{\blambda}{\boldsymbol{\lambda}}
\newcommand{\bbR}{\mathbb{R}}

\newcommand{\cS}{\mathcal{S}}

\newcommand{\EE}{\mathbb{E}}

\newcommand{\Regret}[1]{\operatorname{Regret}\left(#1\right)}
\newcommand{\OPT}{\operatorname{OPT}}

\newcommand{\tA}{\tau_{\pi}}
\newcommand{\pran}[1]{\left(#1\right)}

\newcommand{\ubf}{{\bar{f}}}

\begin{document}
        \author{Yue Xu}
        \email{yuexu.xy@foxmail.com}
        \affiliation{%
        \institution{Alibaba Group.}
        }

        \author{Qijie Shen}
        \email{qijie.sqj@alibaba-inc.com}
        \affiliation{%
        \institution{Alibaba Group.}
        }

        \author{Jianwen Yin}
        \email{yjw264077@alibaba-inc.com}
        \affiliation{%
        \institution{Alibaba Group.}
        }

        \author{Zengde Deng}
        \email{dengzengde@gmail.com}
        \affiliation{%
        \institution{Cainiao Network.}
        }

        \author{Dimin Wang}
        \email{dimin.wdm@alibaba-inc.com}
        \affiliation{%
        \institution{Alibaba Group.}
        }
        
        \author{Hao Chen}
        \email{chenhao@gmail.com}
        \affiliation{\institution{The Hong Kong Polytechnic University.}
        }

        \author{Lixiang Lai}
        \email{lixiang.llx@alibaba-inc.com}
        \affiliation{%
        \institution{Alibaba Group.}
        }

        \author{Tao Zhuang}
        \email{zhuangtao.zt@alibaba-inc.com}
        \affiliation{%
        \institution{Alibaba Group.}
        }

        \author{Junfeng Ge}
        \email{beili.gjf@alibaba-inc.com}
        \affiliation{%
        \institution{Alibaba Group.}
        }


  
	\title{Multi-channel Integrated Recommendation with Exposure Constraints}

	\begin{abstract}
        Integrated recommendation, which aims at jointly recommending heterogeneous items from different channels in a main feed, has been widely applied to various online platforms.
        Though attractive, integrated recommendation requires the ranking methods to migrate from conventional user-item models to the new user-channel-item paradigm in order to better capture users' preferences on both item and channel levels. Moreover, practical feed recommendation systems usually impose exposure constraints on different channels to ensure user experience. This leads to greater difficulty in the joint ranking of heterogeneous items.
        In this paper, we investigate the integrated recommendation task with exposure constraints in practical recommender systems. Our contribution is forth-fold.
        First, we formulate this task as a binary online linear programming problem and propose a two-layer framework named Multi-channel Integrated Recommendation with Exposure Constraints~(MIREC) to obtain the optimal solution.
        Second, we propose an efficient online allocation algorithm to determine the optimal exposure assignment of different channels from a global view of all user requests over the entire time horizon. We prove that this algorithm reaches the optimal point under a regret bound of $ \mathcal{O}(\sqrt{T}) $ with linear complexity.
        Third, we propose a series of collaborative models to determine the optimal layout of heterogeneous items at each user request. The joint modeling of user interests, cross-channel correlation, and page context in our models aligns more with the browsing nature of feed products than existing models.
        Finally, we conduct extensive experiments on both offline datasets and online A/B tests to verify the effectiveness of MIREC.
        The proposed framework has now been implemented on the homepage of Taobao to serve the main traffic, providing service to hundreds of millions of users towards billions of items every day.
	\end{abstract}


	\fancyhead{}
	\settopmatter{printacmref=false,printfolios=false}

	\maketitle

	\section{Introduction}
        Nowadays, the ever-expanding new breeds of content, e.g., pictures, live streams, and short videos, drive the recommender system to drift from the traditional homogeneous form into an integrated form.
        Integrated recommender systems~(IRSs) aim to simultaneously recommend heterogeneous items from multiple sources/channels in a row.
        This integrated form greatly expands users' choices on different types of content thereby satisfying users' diversified preferences on both item-level and channel-level.
        Therefore, IRS has nowadays been widely deployed in various online platforms such as the homepage feeds in Kuaishou~\cite{lin2022feature}, XiaohongShu~\cite{huang2021sliding}, Taobao~\cite{qian2022intelligent}, and AliExpress~\cite{hao2021re}.
        In these products, users continuously slide down to browse and interact with heterogeneous items in a sequential manner, as shown in Figure~1.
        \begin{figure}
        \centering
        \includegraphics[trim = 2 2 2 2, clip, width=0.9\columnwidth]{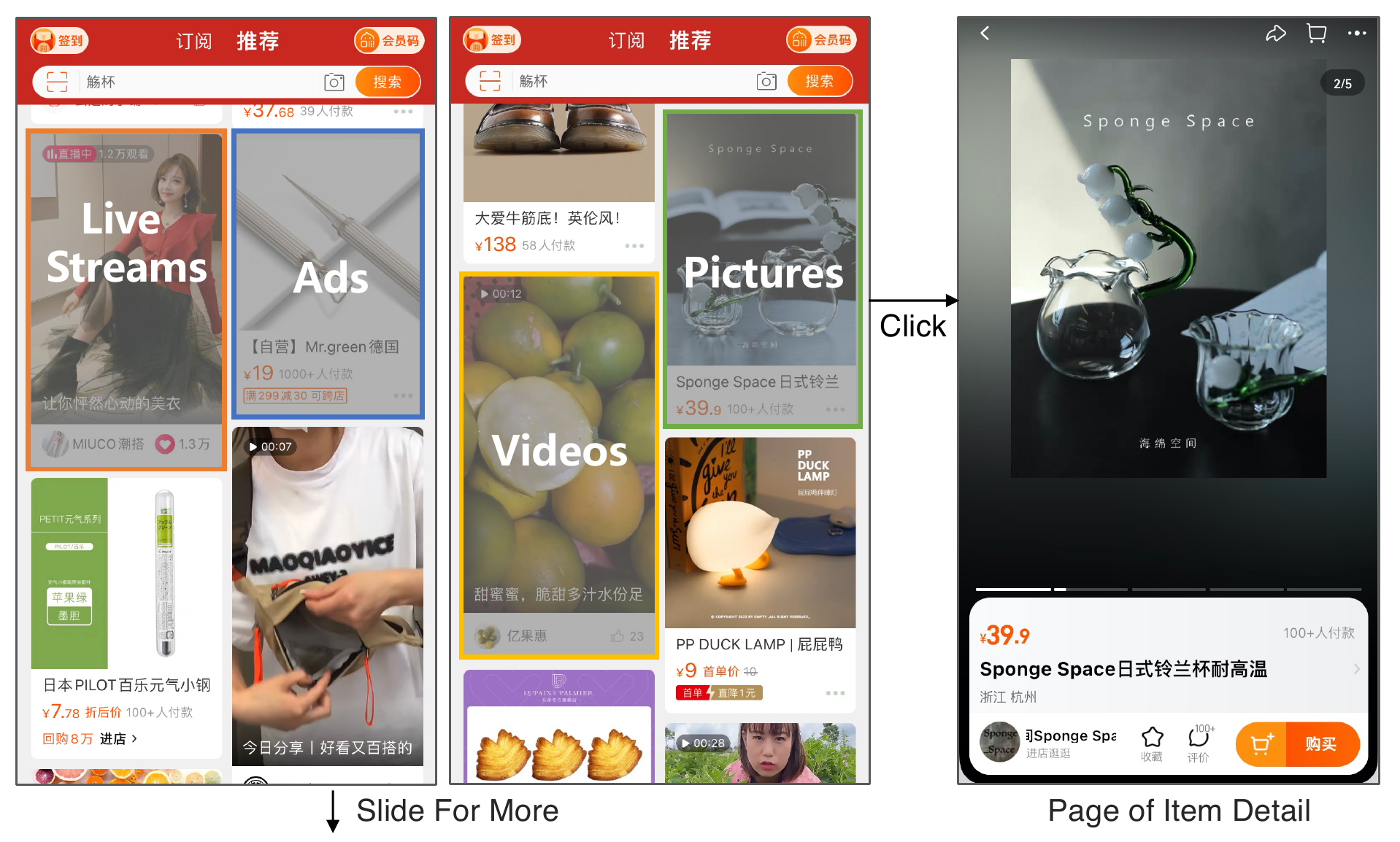}
        \caption{A snapshot of the IRS in real-world feed products. Left: the IRS presents heterogeneous items provided by multiple channels in a row, users slide down to view more items. Right: the detail page is presented after a user clicks an item.}
        \vskip -1.5em
        \label{fig:example}
        \end{figure}

        Though attractive, integrated feed recommendation faces more challenges than conventional recommendation with homogeneous items.
        First, real-world applications usually impose upper or lower exposure guarantees on different channels, such as lower constraints for sponsored/new content~(e.g., ads and cold-start items) or upper constraints for individual channels to ensure diversity. These constraints lead to greater difficulty in the joint ranking of heterogeneous items.
        Second, heterogeneous items from multiple channels usually have different features and ranking strategies. Hence, it is difficult to directly compare items from different channels for joint ranking.
        Third, users' interests on different channels have a great impact on their behaviors, such that traditional user-item prediction models need to evolve into user-channel-item prediction models by considering both intra-channel and inter-channel information and their correlation with user interests.
        Finally, in feed products, users tend to review a large number of items in a row such that the previously viewed items have a great impact on the users' behavior towards the next item~\cite{lin2022feature,huang2021sliding,qian2022intelligent}. Therefore, it is of vital importance to consider the influence from page context when determining the item order in the return list.
        
        Although integrated feed recommendation has been widely deployed in practice, there are still few works focusing on the above challenges systematically.
        In this paper, we propose a general framework named Multi-channel Integrated Recommendation with Exposure Constraints (MIREC) to deal with the multi-channel integrated recommendation task under resource constraints in feed products. 
        %
        MIREC consists of two layers: an allocation-layer which optimizes the exposure of different channels from a global view over all user requests, and a ranking-layer which determines the optimal item layout of a given user request from a local view.
        These two layers operate in an iterative manner to make online decisions along with the arrival of user requests.
	The main contributions are as follows.
	\begin{itemize}
        \item This work formulates the integrated recommendation task with exposure constraints as a binary online linear programming problem and proposes a two-layer framework named MIREC to obtain the optimal solution. We also describe a practical system architecture for its implementation in industrial platforms.
        \item This work proposes an efficient multi-channel allocation algorithm to obtain the optimal exposure assignment of different channels over a fixed time horizon. The proposed algorithm is able to reach an optimal solution with linear complexity w.r.t. the number of constraints. We also prove that this algorithm admits a regret bound of $ \mathcal{O}(\sqrt{T}) $ towards the global optimal point under certain assumptions.
        \item This work proposes a series of collaborative models to determine the optimal layout of heterogeneous items on a page, with joint modeling of user interests, cross-channel correlation, and page context. This aligns more with the browsing nature of feed products than existing models.
        \item This work conducts extensive experiments on both offline datasets and online A/B tests to verify the superiority of our proposed method.
        \end{itemize}
        MIREC has been implemented on the homepage of Taobao to serve the main traffic. It brings $ 3\% $ lift on user clicks, $ 1.56\% $ lift on purchase, and $ 1.42\% $ lift on stay time. It now serves hundreds of millions of users towards billions of items every day. 

	\section{Related Work}
	\smallskip\noindent\textbf{Re-ranking Methods.}
        The main objective of re-ranking methods is to consider the mutual influence among a list of items in order to refine the prediction results produced by point-wise ranking models.
        Three prevalent models are commonly adopted in the existing literature: RNN-based methods, attention-based methods, and evaluator-generator-based methods. The first two methods feed an initial ranking list produced by point-wise models~(e.g., Wide\&Deep~\cite{cheng2016wide}, DIN~\cite{zhou2018deep} and SIM~\cite{pi2019practice}) into RNN-based~(e.g., MiDNN~\cite{zhuang2018globally}, Seq2Slate~\cite{bello2018seq2slate} and DLCM~\cite{ai2018learning}) or attention-based structure~(e.g., PRM~\cite{pei2019personalized}, PFRN~\cite{huang2020personalized}, PEAR~\cite{li2022pear}, and Raiss~\cite{lin2022attention}) sequentially and output the encoded vector at each subsequent layer to model the mutual influences among items. 
        The evaluator-generator-based methods~(e.g., SEG~\cite{wang2019sequential} and GRN~\cite{feng2021grn}), use a generator to generate feasible permutations and use an evaluator to evaluate their list-wise utility to determine the optimal permutation.
        However, most re-ranking methods mainly focus on capturing the mutual influence among homogeneous items provided by one channel, instead of heterogeneous items provided by multiple channels. Moreover, they only optimize the item order at a single time slot, instead of considering a cumulative utility over a broad time horizon under resource constraints.

        \smallskip\noindent\textbf{Online Allocation Methods.}
        The online allocation problem with resource constraints has been mostly studied in online convex optimization~\cite{hazan2016introduction}. 
        The primal-dual methods~\cite{yuan2018online,balseiro2022best,lobos2021joint,chen2023online,yuan2023practice} avoid taking expensive projection iterations by penalizing the violation of constraints through duality.
        The BwK methods~\cite{badanidiyuru2018bandits,immorlica2019adversarial} determines an optimal action from a \textit{finite} set of possible actions and then optimize the policy of decision-making according to the observed rewards and costs over a fixed period of time.
        Several recent works studied the practical performance of online allocation in advertising recommendations.
        For example, PDOA~\cite{zhou2021primal} adopts the primal-dual framework by optimizing the dual prices with online gradient descent to eliminate the online max-min problem’s regret. However, it assumes that the utility and cost values can be optimally estimated and only verify the performance through offline simulations.
        MSBCB~\cite{hao2020dynamic} and HCA2E~\cite{chen2022hierarchically} proposed a two-level optimization framework based on BwK methods, where the high-level determines whether to present ads on the given request while the low-level searches the optimal position to insert ads.
        Most related works on online convex optimization focus on theoretical analysis~(e.g., regret bound) instead of real-world applications. Other related works on advertising mainly consider binary content, i.e., ads or non-ads, instead of heterogeneous content. Directly extending them to deal with multi-channel recommendations in IRSs may lead to sub-optimal results.
    
        \smallskip\noindent\textbf{Integrated Recommendation.}
        The integrated recommendation is a newly emerged but rapidly developing domain driven by practical problems~\cite{liu2022neural}.
        Integrated recommendation methods need to consider both intra-channel and inter-channel features within the heterogeneous content and provide recommendation results continuously along with user arrivals. Recently, DHANR~\cite{hao2021re} proposed a hierarchical self-attention structure to consider the cross-channel interactions. HRL-Rec~\cite{xie2021hierarchical} decomposed the integrated re-ranking problem into two subtasks: source selection and item ranking, and use hierarchical reinforcement learning to solve the problem. DEAR~\cite{zhao2021dear} proposed to interpolate ads and organic items by deep Q-networks. Cross-DQN~\cite{liao2022cross} also adopt a reinforcement learning solution with a cross-channel attention unit.
        However, many integrated methods only focus on ranking at a single time slot instead of over a continuous time horizon. The joint consideration of both integrated ranking and online allocation of limited resources still remains to be explored.

	\section{Problem Formulation}
	This section formulates the integrated recommendation task with exposure constraints as a binary online linear programming problem.
	Specifically, we consider a generic IRS setting where user requests arrive sequentially during a finite time horizon. For each request, the  IRS needs to rank a list of heterogeneous items provided by multiple channels. The aim is to maximize the overall utilities~(e.g., the sum of clicks and pays) of \textit{all channels over the entire time horizon}, subject to multiple resource constraints.

	Formally, the request of user $ u $ triggered at time $ t $ is described as $ e_t = (u, f, g, \mathcal{X}_t) $, where $ f \in \mathbb{R}_{+} $ is a non-negative \textit{utility function},  $ g \in \mathbb{R}_{+} $ is a non-negative resource \textit{consumption function}, and $ \mathcal{X}_t \subset \mathbb{R}^d_+ $ is a compact set denoting all possible item layouts for decision-making.
	For each request $ e_t $, the IRS needs to choose a number of $ N $ heterogeneous items from a candidate set $ I_t $ and place them into $ N $ slots to form a complete page and return it to the user. This action $ \bx_t \in \mathcal{X}_t $ can be represented as a decision matrix $ \bx_t \in [0, 1]^{N \times |I_t| }$, where each entry $ x_{t, n, i} $ is indexed by a slot $ n $ and a card index $ i $.
	Once the user finished viewing the current page, a new user request will be triggered to ask the platform to return to the next page. Therefore, this decision-making process will be performed repeatedly.
	Moreover, in real-world applications, the item layouts need to satisfy the following constraints:
	\begin{equation}
		\mathcal{X}_t  =
		\left\{
		\begin{array}{lr}
			\sum_{i \in I^t} x_{t,n,i} = 1, & \forall t \in \mathcal{T}, \forall n \in \mathcal{N} \\
			\sum_n x_{t,n,i} \leq 1, & \forall t \in \mathcal{T}, \forall i \in I,
		\end{array}
		\right.
	\end{equation}
	where the upper constraint restricts that each slot must be assigned to one item and the lower constraint restricts that each item can be assigned to at most one slot.

	After executing an action $ \bx_t $ at request $ e_t $, the IRS consumes a resource cost $ g(\bx_t) $ and obtains an utility $ f(\bx_t) $.
	In IRS, the utility function $ f(\bx_t) $ is defined according to the concerned metrics. For example, it can be defined as a combination of stay time, adds to cart, and favorites to encourage user engagement, or defined as a combination of clicks and purchases to encourage user conversion.
	On the other hand, the consumption function $ g(\bx_t) $ is defined based on the concerned resource constraints. For example, the platform may need to allocate a certain amount of exposure to new channels in order to support the growth of new content~\cite{gope2017survey}. Meanwhile, a too large proportion of exposures on one specific channel will damage the recommendation diversity thereby harming user experience~\cite{lu2022multi,chen2022hierarchically}. In this case, the IRS needs to guarantee both a lower exposure limit and an upper exposure limit for the heterogeneous items from different channels.

	In this paper, we focus on the exposure constraints in practical systems which lead to the following optimization problem
	\begin{align}\label{P0}
		\mathcal{P}_{0}: \quad  &\text{OPT}(\mathcal{S}) = \max_{\bx_t \in \mathcal{X}_t} \sum\nolimits^T_{t=1} f(\bx_t) \\
		\mathrm{s.t.}  \quad & C_{1}: \sum\nolimits^T_{t=1} g_m(\bx_t) \leq G_{m,th}^{\max} N(\mathcal{S}), \forall m \in \mathcal{M}, \\
		& C_{2}: \sum\nolimits^T_{t=1} g_m(\bx_t) \geq G_{m,th}^{\min} N(\mathcal{S}), \forall m \in \mathcal{M},
	\end{align}
	where $ N(\mathcal{S}) $ denotes the total available exposures to allocate over the entire time horizon, $ G_{m,th}^{\max} $ and $ G_{m,th}^{\min} $ denote the proportion of upper exposure limits and lower exposure limits for each channel $ m \in \mathcal{M} $, respectively, and $ g_m(\bx_t) $ denotes the consumed exposures of cards from channel $ m $ after executing $ \bx_t $ at request $ e_t $.
	Although this paper mainly focuses on the exposure guarantee, the above formulation is generally applicable to other problems with different resource constraints, e.g., the number of coupons to allocate.


	\section{Methodology}
	\begin{figure*}[t]
		\centering
		\includegraphics[trim = 2 2 2 2, clip, width=1.9\columnwidth]{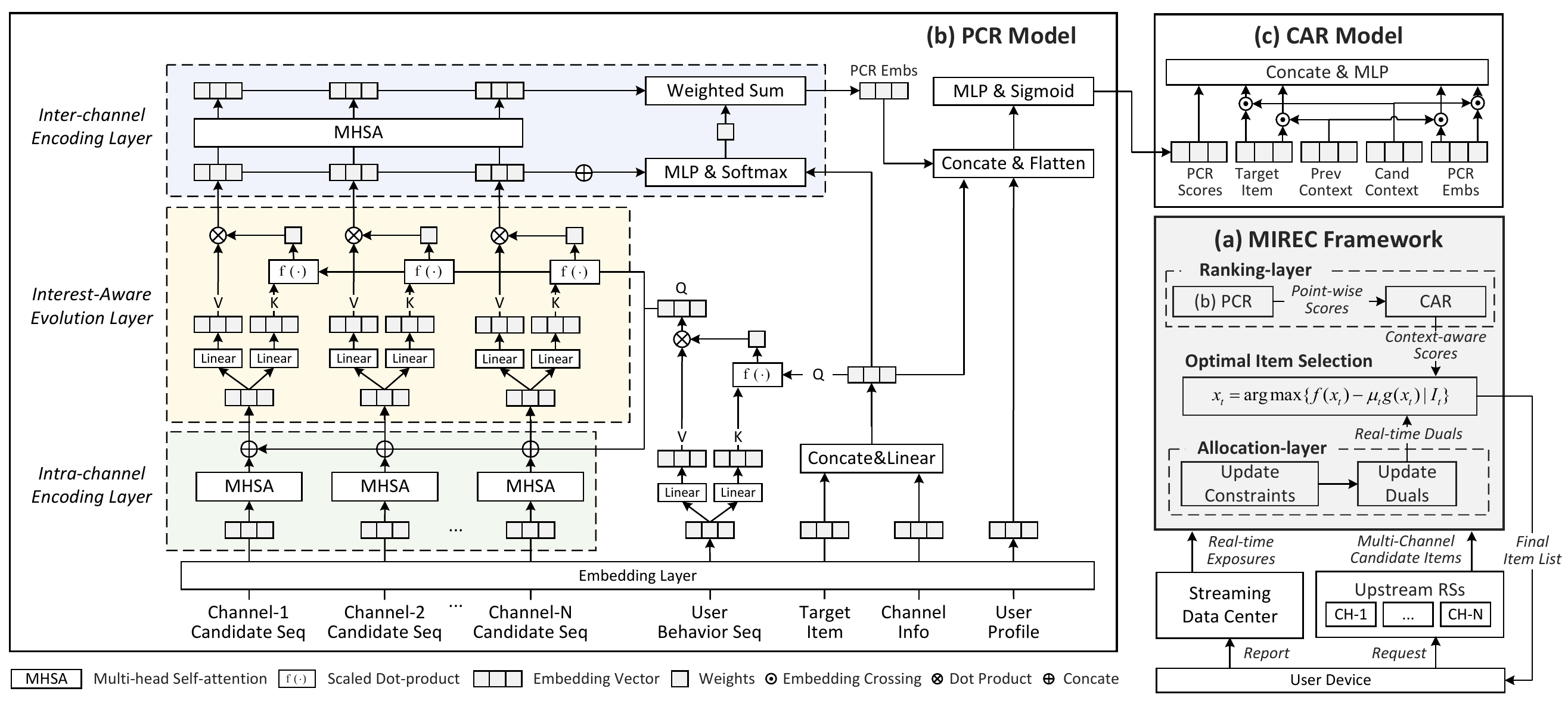}
		\caption{An overview of the MIREC framework.}
		\label{fig:MIREC}
	\end{figure*}

	\subsection{Framework Overview}
	\label{sec:overview}
    Directly solving problem $ \mathcal{P}_0 $  is challenging.
    On one hand, the estimation accuracy of utility $ f(\bx_t) $ and consumption $ g(\bx_t) $ suffer influence from multi-factors, including the user's personal interest, the page context, and the cross-correlation between different channels.
    On the other hand, the determination of each $ \bx_t $ needs to consider the cumulative exposures over the entire time horizon due to the exposure guarantees. Therefore, the optimization of exposure allocation must be performed from a global view over the entire timeline instead of a single time slot.

    To this end, we propose the MIREC framework which solves $ \mathcal{P}_0 $ through online primal-dual iterations. Specifically, MIREC consists of two layers, i.e., the allocation-layer and the ranking-layer, which correspond to the dual and primal problem of $\mathcal{P}_0$, respectively.
    The \textit{allocation-layer} optimizes dual variables to control the cumulative exposure of different channels on all user requests from a global view to guarantee the exposure limits.
    Meanwhile, the \textit{ranking-layer} optimizes the item layout under fixed dual variables given by the allocation-layer, with the aim to maximize the instant utility on a single user request from a local view.
    These two layers operate in an iterative manner along with the arrival of online user requests to determine the optimal item layout at user requests continuously. The general workflow of MIREC is shown in Figure.~\ref{fig:MIREC}.

    For the allocation-layer, we propose a simple but efficient Mirror-descent based Multi-channel Exposure Allocation~(ME2A) algorithm to adaptively balance the utility gain and the exposure cost of presenting heterogeneous items from different channels.
    The proposed M2EA algorithm has a closed form solution that can be computed in linear time and admits a regret bound of $ \mathcal{O}(\sqrt{T}) $ towards the global optimal point under certain assumptions.

    For the ranking-layer, we propose a personalized cross-channel ranking (PCR) model and a context-aware reranking~(CAR) model to jointly determine the optimal item layout on a given user request, with fixed dual parameters given by the allocation-layer. In particular, PCR gives point-wise estimation of the quality of candidate items by joint modeling the influence from user interests, intra-channel information, and inter-channel correlations. Afterward, CAR refines the point-wise estimation generated by PCR into context-aware estimation by making use of both the context information and the high-level knowledge extracted from PCR.

    \subsection{Global: Online Exposure Allocation}
    \label{sec:allocationLayer}
    In this section, we introduce the primal-dual formulation of $ \mathcal{P}_0 $ and propose the ME2A algorithm to obtain the optimal solution for online systems. The complete algorithm is presented in Algorithm~\ref{algorithm}.

    \subsubsection{Primal-dual Formulation}
    The Lagrangian dual function of problem $ \mathcal{P}_{0} $ can be written as
    \begin{align}\label{eq:dual1}
    \min_{\bmu, \blambda} D(\bmu, \blambda) &= \sum_t f(\bx_t) -\sum_m \mu_m \Big(\sum_t g_m(\bx_t)-G_m^{\max}\Big) \\ &+ \sum_m \lambda_m \Big( \sum_t g_m(\bx_t) - G_m^{\min} \Big) \nonumber.
    \end{align}
    Here $ \bmu \geq 0 $ and $ \blambda \geq 0 $ are the introduced dual parameters, $ G_m^{\max} $ and $ G_m^{\min} $ are short for $ G_{m,th}^{\max} N(\mathcal{S}) $ and $ G_{m,th}^{\min} N(\mathcal{S}) $, respectively.
    Note that the parameters $ \bmu $ and $ \blambda $ are related to the violation of exposure consumption over the upper bound limit and the lower bound limit, respectively, which are mutually exclusive.
    In particular, if one of them is positive, the other one must be zero; otherwise, both of them are zero. Hence, it is viable to only introduce one \textit{real number} dual variable $ \bmu \in \mathbb{R}^{1 \times M} $ to replace $ \bmu $ and $\blambda$ in the dual problem, which simplifies~\eqref{eq:dual1} into
	\begin{subequations}\label{eq:dual2}
		\begin{align}
			&\min_{\bmu} D(\bmu) = \sum_t f(\bx_t) - \sum_m[\mu_m]_{+}\Big(\sum_t g_m(\bx_t)-G_m^{\max}\Big)
			\\ &\qquad+\sum_m[-\mu_m]_+ \Big( \sum_t g_m(\bx_t) - G_m^{\min} \Big)  \\
			&= \!\!\sum_t\! \Big( f(\bx_t) \!-\! \!\sum_m\! \mu_m g_m(\bx_t) \Big) \!+\!\!\sum_m\! \Big( [\mu_m]_{+} G_m^{\max} \!-\! [-\mu_m]_+ G_m^{\min}\Big),
		\end{align}
	\end{subequations}
	where $ [\mu_m]_+ = \max\{\mu_m,0\} $.

        \subsubsection{Dual Optimization for Exposure Constraints}
        The dual problem in~\eqref{eq:dual2} can be solved optimally via primal-dual updates.
        Specifically, given a user request $ e_t = (u, f, b, \mathcal{X}_t) $, we assume that the utility $ f(\bx_t) $ and cost $ \bmu_t^T g(\bx_t) $ under different item layout $ \bx_t $ can be properly estimated by the models at the ranking layer. As such, the optimal item layout $ \bx_t $ under a fixed dual variable $ \bmu_t $ can be obtained by solving the following primal problem:
	\begin{equation}\label{eq:primalProblem}
            \mathcal{P}_1: \tilde{\bx}_t = \arg \max_{\bx_t \in \mathcal{X}} \Big\{ f(\bx_t) - \bmu_t^T g(\bx_t) \Big\}.
	\end{equation}
        This is the focus of the ranking-layer to be discussed layer. 
        After the optimal item layout $ \bx_t $ at user request $ e_t $ is properly determined, the next step is to update the dual variable $ \bmu_t $ to adjust the exposure of different channels in future user requests. 
        Specifically, the remained exposure resource of different channels after the presentation of $ \bx_t $ at $ e_t $ is updated by 
	\begin{equation}\label{eq:updateRemainedResource}
		G_{m,t+1}^{\max} = G_{m,t}^{\max} - g_m(\bx_t), \forall m \in \mathcal{M}.
	\end{equation}
	The sub-gradient of the dual function in~\eqref{eq:dual2} can be obtained via Danskin’s theorem~\cite{bertsekas1997nonlinear} by
	\begin{equation}\label{eq:subgradient}
		\nabla \mu_{m,t} = -g_m(\bx_t) + G_{m,t+1}^{\max}\cdot\mathbbm{1}(\mu_{m,t}\geq0) + G_{m}^{\min}\cdot\mathbbm{1}(\mu_{m,t}\leq0),
	\end{equation}
	where $ \mathbbm{1}(x \in A) $ is an indicator function which equals to one if $ x \in A $ otherwise zero.
        As such, the dual variable $ \bmu_t $ can be updated based on the mirror-descent method as
	\begin{equation}\label{eq:mirror_descent}
		\mu_{m,t+1} = \arg\min_{\mu_m \in \mathbb{R}} \mu_{m} \nabla \mu_{m,t}  + \frac{1}{\eta} V_h(\mu_m, \mu_{m,t}),
	\end{equation}
	where $ V_h(x,y) = h(x) h(y) \nabla h(y)^T(x-y) $ is the Bregman divergence based on reference function $ h(\cdot) $ and $ \eta \in \mathbb{R} $ is a fixed step-size.
        Note that this mirror descent step can be computed in linear time since~\eqref{eq:mirror_descent} admits a closed-form solution. For example, if we use $ h(\bmu) = \frac{1}{2}\|\bmu\|^2 $ as the reference function, the dual update in~\eqref{eq:mirror_descent} becomes
	\begin{equation}\label{eq:mirror_descent_simp}
		\bmu_{t+1} = [\bmu_t-\eta \nabla \bmu_{t}]_{+},
	\end{equation}
	which recovers the online projected gradient descent method.
        Moreover, in order to guarantee the upper exposure constraints, one needs to examine the violation of upper limits of each channel before the determination of $ \bx_t $ at each user request. 
        If the sum of exposures of a specific channel exceeds its upper bound, one needs to remove all candidate items from this channel to forbid allocate more exposures when determining $ \bx_t $.
        We present the optimality of this proposed ME2A algorithm and its feasibility to guarantee exposure constraints of different channels as follows. Detailed proofs are deferred to the appendix. 

	\floatname{algorithm}{Algorithm}
	\begin{algorithm}[t]
        \caption{The proposed ME2A algorithm of MIREC}
		\label{algorithm}
		\begin{algorithmic}[1]
			\STATE \textbf{Initialization:}
			\STATE{Initial dual solution $ \mu_1 $, total time periods $ T $, reference function $ h(\cdot) $ and step-size $ \eta $.}
			\STATE \textbf{Iteration:}
			\FOR{$ t = 1, 2, \cdots, T $}
			\STATE{Receive request $ e_t = ( u, f, b, \mathcal{X}_t ) $.}
			\STATE{Update the candidate set $ I_t $ provided by multi-channels.}
			\STATE{Determine the optimal item list $ \bx_t $ by solving the primal problem in~\eqref{eq:primalProblem} at the ranking-layer.
			}
			\STATE{Update the remained exposure resource via~\eqref{eq:updateRemainedResource}.
			}
			\STATE{Obtain the sub-gradient of the dual variable via~\eqref{eq:subgradient}.
			}
			\STATE{Update the dual variable based on mirror descent via~\eqref{eq:mirror_descent_simp}.
			}
			\ENDFOR
		\end{algorithmic}
	\end{algorithm}

	\subsubsection{Optimality}
    It is viable to prove that Algorithm~\ref{algorithm} is asymptotically optimal and admits a regret bound scales as $ \mathcal{O}(\sqrt{T}) $ when the user requests arrive from an \textit{i.i.d} unknown distribution. This assumption is reasonable when the number of requests is numerous~\cite{zhou2021primal,balseiro2022best,lobos2021joint}.
    Specifically, we denote Algorithm~\ref{algorithm} as $ \pi $ and the overall utility over all user requests in set $ \mathcal{S} $ under the running of $ \pi $ as $ R(\pi | \mathcal{S}) = \sum_{t=1}^{T} f(\bx_t). $
	The regret of model $ \pi $ is defined as the worst-case difference over $ \mathcal{S} $ between the expected performance of the global optimal solution and the model $ \pi $:
	\begin{equation}\label{key}
		\text{Regret}(\pi|\mathcal{S}) = \sup \Big\{ \mathbb{E}_{\mathcal{S}}[\text{OPT}(\mathcal{S}) - R(\pi|\mathcal{S})] \Big\},
	\end{equation}
	where $ \text{OPT}(\mathcal{S}) $ denotes the optimal utilities one can obtain under the request set $ \mathcal{S} $.
	The regret bound can be given as follows.
	\begin{Theorem}\label{thm1}
		Suppose that the requests come from an i.i.d model with unknown distribution. Then, $\text{Regret}(\pi|\mathcal{S}) \leq C_1 + C_2 \eta T + \frac{C_3}{\eta}$
		with $ \eta > 0 $ holds for any $ T \geq 1 $. Here $ C_1 $, $ C_2 $ and $ C_3 $ are constant values depending on the numerical bounds of the utility $ f $, the consumption $ g $, and terms from the dual iterates in Eq.~(\ref{eq:mirror_descent}).
	\end{Theorem}
	\noindent From Theorem~\ref{thm1}, we obtain $\mathrm{Regret}(\pi|\mathcal{S}) \leq O(\sqrt{T})$ when using a step-size $\eta \propto c/\sqrt{T}$ with any constant $c>0$. We defer the proof and detailed definitions of $ C_1 $, $ C_2 $, and $ C_3 $ into the appendix.

	\subsubsection{Exposure Feasibility}
    In Algorithm \ref{algorithm}, if the upper exposure limit of a specific channel is violated, we will forbid the exposure of any item from this channel when determining the item list $ \bx_t $. Therefore, the exposure can never be overspent. On the other hand, the lower exposure limits are soft-restricted by adaptively adjusting the dual variable $ \bmu $. This may cause exposure underspent. However, it is viable to prove that the violation of the lower exposure limit of any channel also admits a convergence rate of $ O(\sqrt{T}) $. In other words, even if the violations on lower exposure limits may occur, their growth is considerably smaller than $ T $. 
	\begin{Proposition}\label{prop:lower_bound}
		Suppose the requests come from an i.i.d model with unknown distribution. Then, it holds for any $ T \geq 1 $ and any channel $ m \in \mathcal{M} $ that
		$G_{m}^{\min} - \mathbb{E}\Big[\sum_{t=1}^{T} g_m(\bx_t)\Big]  \le C_4 + \frac{C_5}{\eta}$,
		where $ C_4 $ and $ C_5 $ are constant values depends on the numerical bounds of utility $ f $, consumption $ g $, and terms from the dual iterates~\eqref{eq:mirror_descent}.
	\end{Proposition}
	\noindent Proposition~\ref{prop:lower_bound} states that when using $\eta \propto c/\sqrt{T}$ with $c>0$, the exposure underspend of any channel is bounded by $ O(\sqrt{T}) $. We defer the proof and definitions of $ C_4 $ and $ C_5 $ into the appendix.

	\subsection{Local: Context-Aware Integrated Ranking}
	\label{sec:rankingLayer}
        Different from the allocation-layer which optimizes an objective with accumulative utilities over the entire time horizon as defined in~\eqref{eq:dual2}, the ranking-layer focus on maximizing the utilities on a single time slot.
        This corresponds to the primal problem given in~\eqref{eq:primalProblem}:
        \begin{equation}
        \mathcal{P}_1: \tilde{\bx}_t = \arg \max_{\bx_t \in \mathcal{X}} \Big\{ f(\bx_t) - \bmu_t^T g(\bx_t) \Big\}. \nonumber
        \end{equation}
        In other words, the allocation layer adjusts the exposure of items from different channels by optimizing the dual parameter $ \bmu_t $ from a global view of all user requests. While the ranking-layer determines the optimal item list $ \bx_t $ under a fixed dual variable $ \bmu_t $ from a local view of a given user request $ e_t $. 

    There are two common characteristics that are strongly related to the estimation of $ f(\bx_t) $ and $ g(\bx_t) $ in the integrated recommendation. 
    First, users’ preference on different channels has a great impact on the utility~(e.g., prefer to click or not) and exposure~(e.g., prefer to view or not) estimations. Therefore, it is of vital importance to consider both intra-channel and inter-channel correlations with reference to user interests during the estimation. 
    Second, in feed products, users tend to review a large number of items in a row such that the previously viewed items have a great impact on users’ behavior towards the next item. Therefore, it is necessary to consider page context when determining the item order.

    Therefore, we propose two models to deal with the above two challenges, respectively. 
    First, we propose PCR model to deal with the joint modeling of user interests and inter/intra-channel correlation. It gives a point-wise estimation of the utility/exposure value of presenting each candidate item.
    Second, we propose CAR model to refine the point-wise estimation from PCR into context-aware estimation by considering both context information and the high-level knowledge obtained from PCR. 
    It is also responsible for selecting optimal items from a set of candidate items to generate the final return list. 
    In real-world systems, for each user request, we only need to run PCR once to get the point-wise scores, and then run CAR multiple times to generate the return list. 
    Next, we mainly focus on the estimation of $ f(x) $, the estimation of $ g(x) $ can be performed in a similar way by changing the learning goals.




	\subsubsection{Personalized Cross-Channel Ranking Model.}
	PCR takes four types of features as input, i.e., the user profile feature $X_{u}$, the user behavior sequences $X_{b}$, the candidate items provided by each channel $X_{l}$, and the item-level features of target item $X_{i}$. As shown in Figure~\ref{fig:MIREC}, We use an embedding layer to transform these features into dense embedding vectors, denoted as $E_{u}$, $E_{b}$, $E_{l}$ and $E_{i}$, respectively.
	These embedding vectors are then fed into three components, i.e., the intra-channel encoding layer, interest-aware evolution layer, and inter-channel encoding layer in order, which are described below.

	\smallskip\noindent\textbf{Intra-Channel Encoding layer.}
	This layer aims at extracting the mutual influence of item pairs and other extra information within the channel. We adopt the well-known multi-head attention~\cite{vaswani2017attention} as the basic learning unit for intra-channel encoding. This is due to that the self-attention mechanism is able to directly capture the mutual influences between any two items, and is robust to far distance within the sequence.
	Formally, the formulation of this attention-based encoding can be written as
	\begin{subequations}
		\begin{align}
			V_{l}^{m}&=[head_{1},head_{2},...,head_{h}]W^{O}, \label{eq:lstm_form5} \\
			head_{i} &= \text{Softmax}\left(\frac{(E_{b}W_{Q})(E_{b}W_{K})^{T}}{\sqrt{d_{h}/h}}\right)\left(E_{b}W_{V}\right), \label{eq:form5}
		\end{align}
	\end{subequations}
	where $W^{O}\in \mathbb{R} ^{d_{h}\times d_{h}} $ denotes the learnable parameters for each head with $ d_h $ being the length of projected embedding vector after attention, $W_{Q}, W_{K}, W_{V}\in \mathbb{R}^{d\times d_{h}/h}$ are the vectors of query, key and value with $ d $ being the length of original embedding vector and $h$ being the number of heads,  $V_{l}^{m}$ represents the encoded candidate items of each channel $ m \in \mathcal{M} $.

	\smallskip\noindent\textbf{Interest-Aware Evolution Layer.}
	Existing works such as PRM~\cite{pei2019personalized} and DHANR~\cite{hao2021re} directly apply the self-attention mechanism to model the inter-dependencies among items and channels without considering user's recent behavior. However, the interests hidden in user's behavior items usually have a great impact on the prediction accuracy in recommendation tasks~\cite{zhou2018deep,zhou2019deep,pi2019practice,chen2022efficient}.
	The recently proposed PEAR~\cite{li2022pear} firstly models the dependency between the candidate item list and the user's historical behaviors based on a transformer-like structure, which, however, suffers from two limitations.
	First, directly mixing the raw item-level features from user behavior items may introduce redundant or noisy information to degrade the learning performance.
	Second, each user may exhibit multiple interest points, such that it is beneficial to reinforce the interest related to the target item before feature-crossing to avoid drifting.
	Therefore, we first reinforce the interest vector according to the correlation between behavior items and the target item as
        \begin{equation}
            V_{U} =f(E_{b};E_{i})=\sum\nolimits_{i=1}^{B}A(b_{i},E_{i})b_{i}= \sum\nolimits_{i=1}^{B}w_{i}b_{i},
        \end{equation}
	where $B$ is the length of behavior sequence, $b_i$ is the behavior item, $V_{U}$ denotes the user representation feature with respect to $E_{i}$, and $A(\cdot)$ is a feed-forward network whose output is the activation weight $w_{i}$.
	Then, we make use of this reinforced interest vector to extract useful information from the candidate items of different channels.
	Formally, for each channel $m$, given $V_{l}^{m}$ and $V_{U}$ as inputs, we use scaled dot-product attention formulated as follows:
	\begin{equation}
		H_{s}^{i}=\text{Softmax}\left(\frac{(V_{l}^{m}W_{q})[V_{U}W_{k1},V_{l}^{m}W_{k2}]^{T}}{\sqrt{d_{h}}}\right)[V_{U}W_{v1},V_{l}^{m}W_{v2}],
	\end{equation}
	where $W_{k1},W_{v1}\in \mathbb{R}^{d\times d_{h}}$ and $W_{k2},W_{q}, W_{v2} \in \mathbb{R}^{d_{h}\times d_{h}}$ are all learnable parameters, $[ \cdot ]$ denotes the concatenation operation.
	After the above operations, we successfully merged the information from the candidate item lists and the user's historical behaviors into a series of evolved embedding vectors for further processing.

        \smallskip\noindent\textbf{Inter-Channel Encoding Layer.}
        Previous layers mainly extract intra-channel correlations. We now focus on the modeling of inter-channel correlation.
        First, we feed the embedding vector of each channel and the target item embedding into the MLP layer with softmax function to obtain the importance weights on each channel that is related to the target item: 
        \begin{equation}
         W_{CH} = \text{Softmax}(\text{MLP}[H_{s}^{m},E_{i}]),
        \end{equation}
        where $ W_{CH} \in \mathbb{R}^{1\times m} $ is the importance weights, and $H_{s}^{m}$ denotes the concatenation of all channels' output from the Interest-Aware Evolution Layer.
        Second, we perform multi-head self-attention on the evolved embedding vector $H_{s}^{m}$  of each channel $m~\in~\mathcal{M}$ to obtain the mixed embedding $\tilde{H}_{s}^{m}$ which contains rich inter-channel information.
        Then, we perform the weighted sum on the mixed embedding of all channels based on $ W_{CH} $ to get the final representation of multi-channel modeling: 
        \begin{equation}
        V_{L} = W_{CH} \cdot [\tilde{H}_{s}^{m}]^T, m \in \mathcal{M},
        \end{equation}
        where $ [\tilde{H}_{s}^{m}] $ represents the concatenation of the mixed embeddings of all channels.
        

        Finally, we concatenate all vectors as input and feed it into the MLP layers with a sigmoid function to predict the utility of presenting a given target item to a given target user as
        \begin{equation}
        \begin{split}
            Y_{PCR} = \text{Sigmoid}(\text{Concat}(E_{u},E_{i},V_{L}))
        \end{split}
        \end{equation}

        \subsubsection{Context-Aware Refinement Model.}
        In this section, we propose the CAR model to refine the point-wise utility scores estimated by PCR into context-aware utility scores.
        Given a candidate item set $I_{cand}$ with size $ N $, the aim of CAR is to optimally choose $ K $ items from $I_{cand}$ and allocate them to the $ K $ slots in a page based on the learning results from PCR.

        We maintain two types of context information in CAR, i.e., the context of previous items and the context of remaining candidate items.
        Specifically, when selecting the $ k $-th item in a page, we represent the context of previously presented $ k-1 $ items $h_{pre}$ by mean-pooling over their embeddings. Meanwhile, we represent the context of all candidate items $h_{can}$ by mean-pooling over the embeddings of all remained candidates. These two context vectors are updated and repeated along with the item selection process.
        Furthermore, we perform a series of embedding crossing operations between the target item embedding $ e_i $ and the context embeddings to model the influence from page context.
        In specific, the operations in the context of previous items can be formulated as follows:
        \begin{equation}\label{eq:prevContextAgg}
            H_{pt} = \text{Concat}(h_{pre} \oplus e_i, h_{pre} \otimes e_i, h_{pre} \ominus e_i),
        \end{equation}
        where $ \oplus $, $ \otimes$, and $ \ominus $ denote the addition, subtraction, and dot product between embedding vectors, respectively.
        The same goes for $H_{ct}$ by replacing $ h_{pre} $ in~\eqref{eq:prevContextAgg} with the context of candidate items $h_{can}$.
        Additionally, we also perform embedding-crossing between the context embeddings and the high-level knowledge $ V_L $ from PCR to obtain another two vectors, i.e., $H^{v}_{ct}$ and $H^{v}_{pre}$.

        Finally, for each candidate item $ i \in I_{cand} $, we predict the context-aware utility score by feeding these embedding vectors along with the point-wise utility score $Y_{PCR}$ from PCR and user profile features $E_{u}$ into one MLP layer as
	\begin{subequations}
		\begin{align}
            H_{\text{all}} &\!=\! \text{Concat}(H_{pt},H_{ct},H^{v}_{ct},H^{v}_{pre},h_{pre},h_{can},E_{u},Y_{PCR}),\\
            Y_{CAR} &\!=\! \sigma(\text{MLP}(H_{\text{all}})),
		\end{align}
	\end{subequations}
        where $\sigma$ represents the sigmoid activation function.
        After scoring all candidate items, we choose the item with the highest score as the optimal item for slot $ k $, and update the context vectors and the remained candidate items accordingly. This process will be repeated $ K $ times to generate a return item list of length $ K $. Note that the above operations in CAR only involve linear computations, such that this item selection process is cost-efficient in online systems.
        
        Both PCR and CAR can be trained with the commonly used cross-entropy loss as in other ranking models, the learning objective       can be given as
        \begin{equation}\label{eq:loss}
        J = \sum\nolimits_{e_t \in \mathcal{D}} \Big( y^{e_t}_{u,i} \log \hat{y}^{e_t}_{u,i} + (1-y^{e_t}_{u,i}) \log (1-\hat{y}^{e_t}_{u,i}) \Big),
        \end{equation}
        where $ \mathcal{D} $ denotes the training dataset, $ y^{e_t}_{u,i} $ is the real user-item recommendation label (equals $ 1 $ or $ 0 $) between user $u$ and item $i$ at request $e_t$, and $ \hat{y}^{e_t}_{u,i} $ is the predicted label given by $Y_{PCR}$ or $Y_{CAR}$. 
        In our experiments, when predicting the utility function $f$ with user clicks, $ y^{e_t}_{u,i} $ refers to the click label; when predicting the cost function $g$ with exposure constraints,  $ y^{e_t}_{u,i} $ refers to the  exposure label between user $u$ and item $i$ at request $e_t$, i.e., whether user $u$ has seen item $i$ at request $e_t$.
        One can readily change the learning objective according to actual demands.

	\subsection{Online Implementation}
	\label{sec:implementation}
	\begin{figure}
		\centering
		\includegraphics[trim = 2 2 2 2, clip, width=0.75\columnwidth]{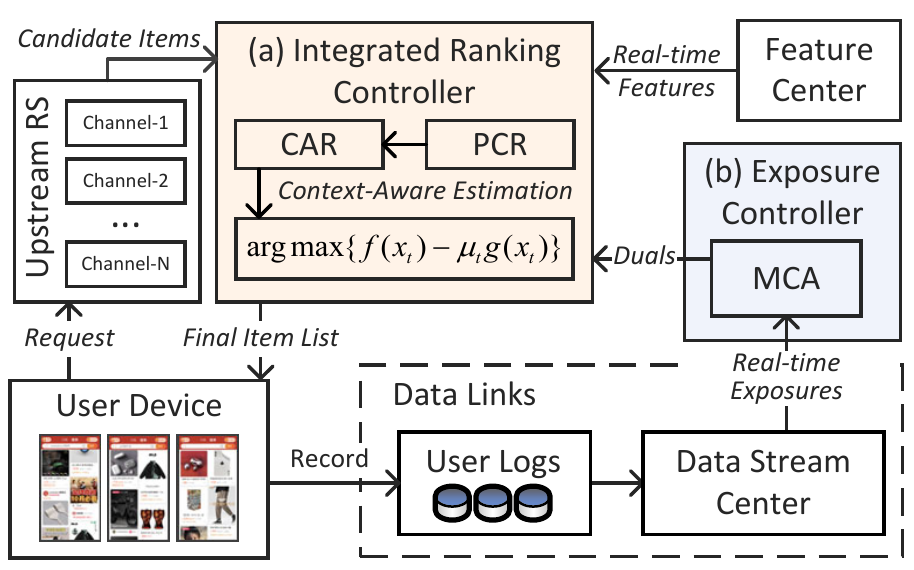}
		\caption{Online system architecture.}
		\label{fig:implementation}
	\end{figure}
        In this section, we introduce the online implementation of our proposed MIREC model in the homepage feed of Taobao. The presented system architecture is able to handle $ 120, 000 $ QPS at traffic peak and respond within $ 20 $ milliseconds in general. It now serves the main traffic of Taobao to provide services to hundreds of millions of users towards billions of items in Taobao every day.
        
        Figure.~\ref{fig:implementation} gives a general architecture to implement our proposed MIREC model in real-world IRS. Each time a user request is triggered from the device, the upstream RS of each channel will run its own recommendation models to determine the top items to return. The Integrated Recommendation Controller uses the top items from all channels as the candidates. It retrieves user/item features from a feature center in real-time and ranks candidate items by solving based on our proposed LCPR and PCR model. Meanwhile, the dual variable $ \mu $ is estimated by an Exposure Controller. This module monitors the completeness of exposure guarantees based on the real-time exposures collected from user logs and updates the dual variable to adjust the exposures on different channels periodically.

	\section{Experimental Results}
        This section conducts extensive experiments on both offline datasets and real-world applications with the goal of answering the following research questions:
        
        \smallskip\noindent\textbf{Q1:} Does our proposed PCR and CAR outperform other baseline models in integrated ranking tasks?
        
        \smallskip\noindent\textbf{Q2:} Does our proposed MIREC framework outperform other methods in integrated recommendation tasks with exposure constraints?
        
        \smallskip\noindent\textbf{Q3:} How does MIREC perform in real-world applications?
        
	\subsection{Experimental Setup}
	\subsubsection{Datasets}
        We use one public dataset named MicroVideo-1.7M and one industrial dataset named Taobao for experiments.
        The public available MicroVideo-1.7M dataset\footnote{\url{https://github.com/Ocxs/THACIL}} released by~\cite{chen2018thacil} contains $ 12,737,619 $ interactions that $ 10,986 $ users have made on $ 1,704,880$ micro-videos. This dataset provides rich user behavior data and timestamps to evaluate the performance on both interest modeling and context-aware reranking.
        %
        The Taobao dataset is an industrial private dataset that contains users' behaviors and feedback logs from multiple channels in the homepage feed of Taobao Mobile App. It is one of the largest feed scenarios for online merchandise in China.
        The feed provides items in form of the streams, videos, pictures, etc, from various channels. Users can slide to view more items in a row.
        This dataset contains about ten billion interactions that one hundred million of users have made on sixty million items.
        We also conduct online A/B tests on the platform Taobao to examine the performance of MIREC in real-world applications.

	\subsubsection{Comparing Methods}
        We compare MIREC with two mainstreams of baselines.
        The first steam of baselines are the methods for ranking tasks with different goals on user interest modeling~(i.e., DIN and DIEN), re-ranking~(i.e., DLCM, PRM, and PEAR), or multi-channel recommendation~(i.e., STAR and DHANR).
        Specifically, \textbf{DIN}~\cite{zhou2018deep} is a widely used benchmark for sequential user data modeling in point-wise CTR predictions, which models short behavior sequences with target attention.
        \textbf{DIEN}~\cite{zhou2018deep} combines GRUs and attention to capture temporal interests from users' historical behaviors with respect to the target item.
        \textbf{DLCM}~\cite{ai2018learning} uses gated recurrent units (GRU) to sequentially encode the top-ranked items with their feature vectors.
        \textbf{PRM}~\cite{pei2019personalized}: directly optimizes the whole recommendation list by employing a Transformer structure to efficiently encode the information of all items in the list.
        \textbf{PEAR}~\cite{li2022pear} not only captures feature-level and item-level interactions but also models item contexts from both the candidate list and the historical clicked item list.
        \textbf{STAR}~\cite{sheng2021one} trains a unified model to serve all channels simultaneously, which consists of shared centered parameters and channel-specific parameters.
        \textbf{DHANR}~\cite{hao2021re} proposes a hierarchical self-attention structure to consider cross-channel interactions.

        The second stream of baselines is the online allocation methods which have been successfully applied in industrial applications for online resource allocation.
        \textbf{Fixed} is the fixed-positions strategy, where the positions of recommended items and ads are manually pre-determined for every request.
        \textbf{$\beta$-WPO}: is based on the Whole-Page Optimization (WPO)~\cite{zhang2018whole}. WPO ranks recommended and ad candidates jointly according to the predefined ranking scores. Similar to~\cite {chen2022hierarchically}, we introduce an adjustable variable $ \beta $ to control the proportion of different channels on each request to satisfy the resource constraint. In general, $\beta$-WPO can be regarded as a heuristic list merging algorithm. Each list from one channel is assigned a priority weight. The algorithm merges the top items of each list based on both their ranking scores and the priority weights into a final return list.
        \textbf{HCA2E}~\cite{chen2022hierarchically}: proposed a two-level optimization framework based on BwK methods. The high-level determines whether to present ads on the page while the low-level searches the optimal position to insert ads heuristically.
	
	\subsubsection{Metrics.}
        For offline experiments, we use user clicks to measure the utility function $f$. For online experiments, we consider a joint measurement of user click, purchase, and stay-time for utility function $f$. For all experiments, we use the exposure of items to measure cost function $g$.
	In this case, we compare the performance of integrated ranking in offline evaluation using the widely used Area Under ROC (AUC) and normalized discounted cumulative gain (nDCG)~\cite{jarvelin2017ir}.
	Remark that nDCG@K refers to the performance of top-k recommended items in the return list.
	The online performance is evaluated by CLICK, Click-Through-Rate~(CTR), Gross Merchandise Volume~(GMV), and Stay Time.
	Here, CLICK refers to the total number of clicked items. CTR is defined as \mbox{CLICK/PV} with PV denoting the total number of impressed items. CTR measures users' willingness to click and is therefore a widely used metric in practical applications. GMV is a term used in online retailing to indicate a total sales monetary-value for merchandise sold over a certain period of time.
	Stay Time denotes the time period of users' average stay time in the product, averaged on all users.

	\subsubsection{Reproducibility.}
	\label{sec:parameter_setting} 
        Our source codes have been made public to ensure reproducibility\footnote{\url{https://github.com/anonymousauthor7/MIREC}}.
	In all experiments, we use the validation set to tune the hyper-parameters to generate the best performance for different methods. The learning rate is searched from $10^{-4}$ to $10^{-2}$. The L2 regularization term is searched from $10^{-4}$ to $1$. All models use Adam as the optimizer.
	\begin{table}[]
		\caption{Comparison of ranking performance~(bold: best; underline: runner-up).}
		\resizebox{0.9\columnwidth}{!}{
			\begin{tabular}{@{}cccccc@{}}
				\toprule
				Dataset                     & Method & AUC             & Logloss         & NDCG@20         & NDCG@30         \\ \midrule
				\multirow{8}{*}{MicroVideo\tablefootnote{Note that MicroVideo is a public dataset with a single channel, such that we omit the comparison with STAR and DHANR which are proposed for multi-channel modeling. We will release the multi-channel dataset collected in Taobao for future research.}} & DIN    & 0.6831          & 0.5922          & 0.5403          & 0.6535          \\
				& DIEN   & 0.6842          & 0.5909          & 0.5408          & 0.6537          \\ \cmidrule(l){2-6}
				& DLCM   & 0.6872          & 0.5898          & 0.5582          & 0.6698          \\
				& PRM    & 0.6979          & 0.5872          & 0.5591          & 0.6708          \\
				& PEAR   & {\ul 0.7021}    & {\ul 0.5821}    & {\ul 0.5632}    & {\ul 0.6745}    \\ \cmidrule(l){2-6}
				
				& Ours   & \textbf{0.7084} & \textbf{0.5787} & \textbf{0.5667} & \textbf{0.6826} \\ \midrule
				\multirow{8}{*}{Taobao}     & DIN    & 0.7681          & 0.4982          & 0.5203          & 0.6481          \\
				& DIEN   & 0.7692          & 0.4971          & 0.5202          & 0.6479          \\ \cmidrule(l){2-6}
				& DLCM   & 0.7699          & 0.4965          & 0.5209          & 0.6482          \\
				& PRM    & 0.7722          & 0.4941          & 0.5211          & 0.6489          \\
				& PEAR   & 0.7748          & {\ul 0.4919}    & 0.5232          & 0.6511          \\ \cmidrule(l){2-6}
				& STAR   & 0.7738          & 0.4931          & 0.5219          & 0.6492          \\
				& DHANR  & {\ul 0.7753}    & 0.4923          & {\ul 0.5243}    & {\ul 0.6513}    \\ \cmidrule(l){2-6}
				& Ours   & \textbf{0.7791} & \textbf{0.4899} & \textbf{0.5275} & \textbf{0.6545} \\ \bottomrule
			\end{tabular}
		}
		\label{tab:itemQualityMeasurement}
        \vskip -1em
	\end{table}

	\begin{table}[]
		\small
		\caption{Ablation study of the ranking components.}
		\resizebox{1.0\columnwidth}{!}{
			\begin{tabular}{l|cccc}
				\toprule
				& AUC      & Logloss & {NDCG@20} & {NDCG@30} \\ \midrule
				PCR$^{*}$             & 0.7758 & 0.4933  & 0.5222                               & 0.6511                               \\
				PCR$^{\dagger}$              & 0.7761 & 0.4932  & 0.5246                               & 0.6513                               \\
				PCR w/o IntraCE          & 0.7763 & 0.4928  & 0.5247                               & 0.6516                               \\
				PCR w/o InterCE          & 0.7756 & 0.4935  & 0.5239                               & 0.6509                               \\
				PCR                      & 0.7778   & 0.4913  & 0.5262                               & 0.6531                               \\
				PCR+CAR (propsed) & 0.7791   & 0.4899  & 0.5275                               & 0.6545         \\ \bottomrule
			\end{tabular}
		}
		\vskip -1em
		\label{tab:ablationStudy}
	\end{table}

	\subsection{Offline Evaluation}
	\subsubsection{Q1: Performance on Integrated Ranking}
	We first compare the performance with the first stream of baselines on item ranking.
	The results are shown in Table~\ref{tab:itemQualityMeasurement}, which leads to the following findings.
	First, the re-ranking methods perform generally better than the point-wise user interest methods, indicating that modeling mutual influence among the input ranking list is of vital importance for the ranking. Therefore, it is essential to consider the influence from page context in feed recommendations.
	Second, the multi-channel methods perform better than the reranking methods, which verifies that exploiting the distinction and mutual influence among different channels has a great impact on integrated recommendations.
	Besides, we also notice that DHANR performs better than STAR, which may be due to that DHANR considers both the correlation among different channels and the influence from the candidate list.
	Finally, our proposed MIREC model achieves superior performance than other competitors on all datasets, verifying the effectiveness of joint modeling the cross-channel information, user interest, context information, and candidate list.

	\subsubsection{Ablation Study}
	The results in Table.~\ref{tab:ablationStudy} investigates the impact of each component of MIREC on item quality estimation.
	Specifically, PCR$^{*}$ replaces the attention mechanism for user behaviors in the Merged-Sequence Evolution layer with a self-attention mechanism which is in accordance with PEAR~\cite{li2022pear}. PCR outperforms PCR$^{*}$ indicates that the tailored attention mechanism in PCR can filter out noisy or redundant information from historical behaviors to benefit the subsequent modeling of bi-sequence interaction. PCR$^{\dagger}$ removes the scaled dot-product attention mechanism (i.e. there is no explicit interaction between initial lists and user behaviors) and achieved a worse performance. This demonstrates the necessity of this direct modeling between sequences, directly guiding the reordering of the initial lists.
	PCR w/o IntraCE removes the Intra-Channel Encoding module, i.e., directly feeding the embeddings of the initial item lists into subsequent layers for learning. The result shows that PCR achieves superior performance than PCR w/o IntraCE, verifying that it is of vital importance to model the mutual information inside each channel for final prediction.
	PCR w/o InterCE removes the Inter-Channel Encoding, which also leads to worse performance.
	It verifies that without considering the relationship and distinction between different channels will degrade the model performance considerably.
	Moreover, the joint learning of PCR and CAR performs better than only using PCR.
	This verifies that the modeling of page-wise context information can improve prediction accuracy effectively.

	\begin{table}[]
		\centering
		\caption{Joint performance of allocation and ranking~(bold: best; dagger: baseline).}
		\resizebox{0.95\columnwidth}{!}{
			\begin{tabular}{@{}cccccccc@{}}
				\toprule
				\multirow{2}{*}{Exp. Settings} & \multirow{2}{*}{Method} & \multicolumn{4}{c}{Exposure completeness} & \multirow{2}{*}{CTR} & \multirow{2}{*}{CTR Lift} \\ \cmidrule(lr){3-6}
				&       & CH1    & CH2    & CH3    & CH4    &                 &         \\ \midrule
				\multirow{4}{*}{Setting-1} & Fixed & 0.44\% & 1.35\% & 0.20\% & 0.50\% & $ 5.54\%^{\dagger} $    & -       \\
				& WPO   & 0.15\% & 0.15\% & 0.13\% & 0.30\% & 6.09\%          & +9.93\%  \\
				& HCA2E & 0.24\% & 0.95\% & 0.33\% & 0.10\% & 6.34\%          & +14.44\% \\
				& Ours  & 0.02\% & 0.65\% & 0.67\% & 0.20\% & $ \textbf{6.56\%}^*$ & $ \textbf{+18.41\%}^*$ \\ \midrule
				\multirow{4}{*}{Setting-2} & Fixed & 0.10\% & 0.53\% & 0.10\% & 0.20\% & $5.91\%^{\dagger}$    & -       \\
				& WPO   & 0.16\% & 0.27\% & 1.40\% & 0.20\% & 6.28\%          & +6.26\%  \\
				& HCA2E & 0.19\% & 0.53\% & 0.30\% & 0.40\% & 6.53\%          & +10.49\% \\
				& Ours  & 0.17\% & 0.53\% & 0.40\% & 0.20\% & $ \textbf{6.76\%}^*$ & $ \textbf{+14.38\%}^* $ \\ \bottomrule
		\end{tabular}}
		\label{tab:jointPerformance}
	\end{table}

	\subsubsection{Q2: Performance with Exposure Constraints}
	To the best of our knowledge, there does not exist publicly available datasets which has rich user logs and multi-channel features to examine the joint performance of integrated recommendation and exposure allocation. Therefore, we only perform experiments on Taobao dataset, using the complete platform logs. In this experiment, we assume that the IRS needs to allocate exposures to satisfy the exposure guarantees of four distinct channels. The aim is to maximize the overall user-click utility of all channels. The compared fixed, WPO, and HCA2E methods all use point-wise scores to be consistent with their original proposals. For HCA2E, we use their proposed heuristic search method to determine the final order of the item list.
	The results are shown in Table~\ref{tab:jointPerformance}, which are averaged on multiple runs to give a fair comparison. The simulated time horizon is one complete day with more than one billion user requests from real productive environment.
	We evaluate the performance using two different sets of lower bounds: 1) Channel~1=55\%, Channel~2 = 20\%, Channel~3 = 15\%, Channel~4 = 10\%; 2) Channel~1=70\%, Channel~2 = 15\%, Channel~3 = 10\%, Channel~4 = 5\%. The parameters of all comparing methods are carefully tuned to satisfy the exposure constraints. The completeness in Table~\ref{tab:jointPerformance} shows that all methods can control the violation of constraints to a low-level. HCA2E and our proposed MIREC perform slightly better than the fixed method and the WPO method.
	Noticeably, our proposed method outperforms other comparing methods considerably in terms of CTR enhancement, which verifies that the joint use of the allocation and estimation algorithm can bring a remarkable improvement in practical environments.

	\subsection{Online Evaluation}
	\begin{table}[]
		\centering
		\caption{Results of online A/B tests. }
		\scalebox{0.9}{
			\begin{tabular}{@{}ccccccc@{}}
				\toprule
				& CLICK & CTR & GMV    & Stay Time    \\ \midrule
				Ours vs Fixed    & +4.02\% & +2.15\%    & +1.98\%    & +2.01\% \\
				Ours vs Baseline    & +3.00\% & +1.75\%    & +1.56\%    & +1.42\% \\ \bottomrule
			\end{tabular}
		}
            \vskip -1em
        \label{tab:online}
	\end{table}
	\begin{figure}[t]
		\centering
		\subfigure[Comparison of the stability of exposure propotion.]
		{
			\label{fig:dailyPVR}
			\includegraphics[trim = 2 2 2 2, clip, width=0.6\columnwidth]{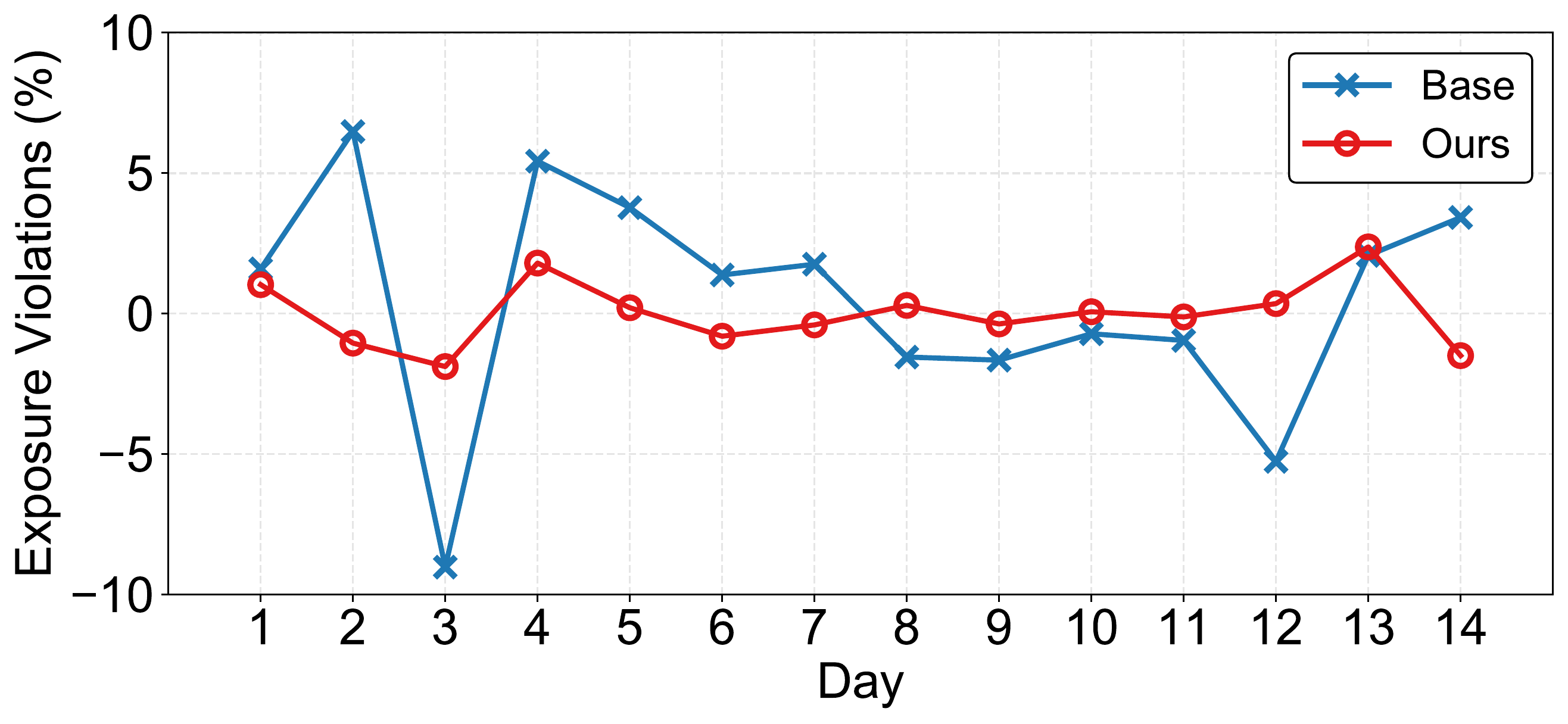}
		}
		\subfigure[Exposure distribution.]
		{
			\label{fig:posPVR}
			\includegraphics[trim = 2 2 2 2, clip, width=0.365\columnwidth]{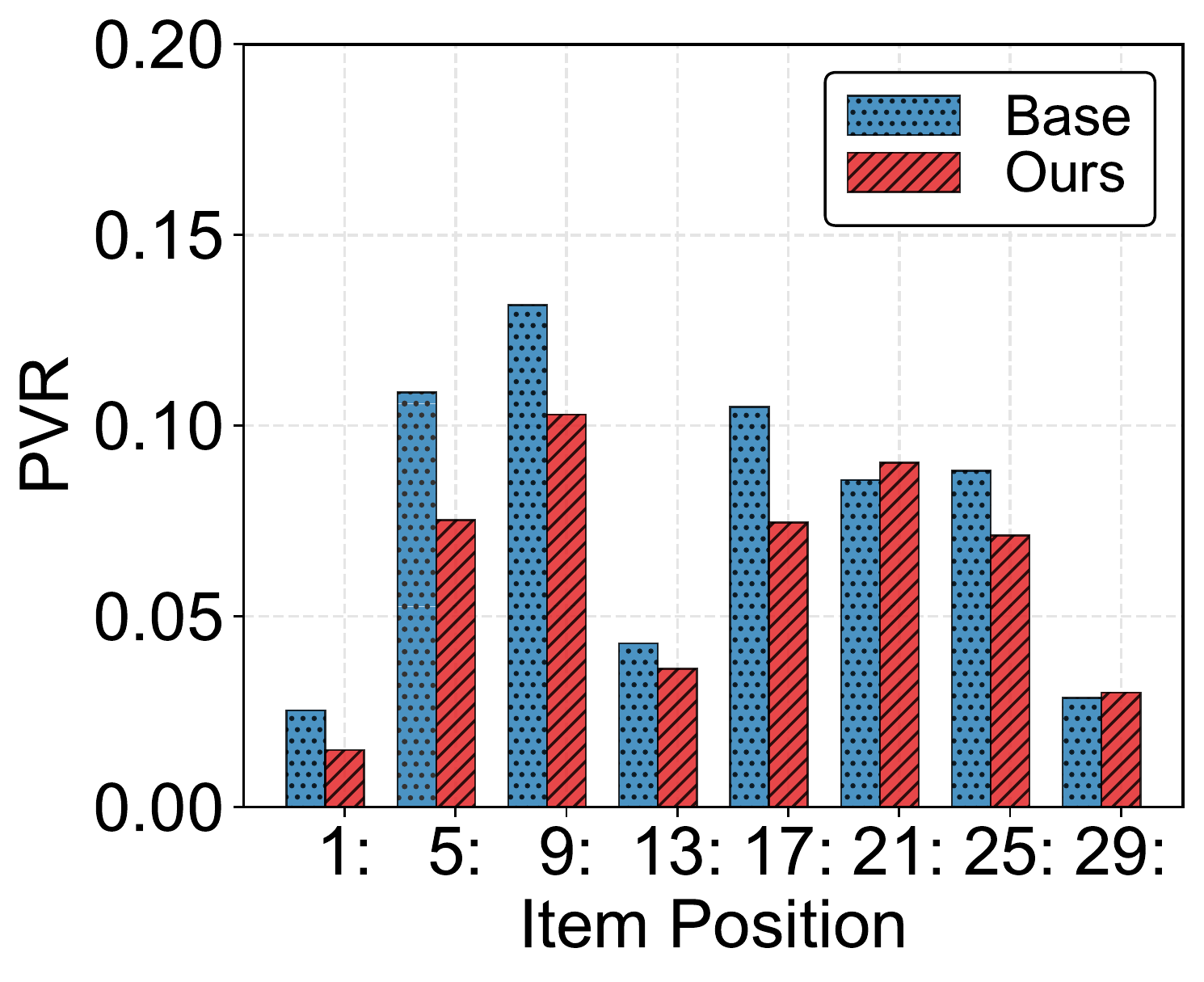}
		}
		\subfigure[CTR on different positions.]
		{
			\label{fig:posCTR}
			\includegraphics[trim = 2 2 2 2, clip, width=0.35\columnwidth]{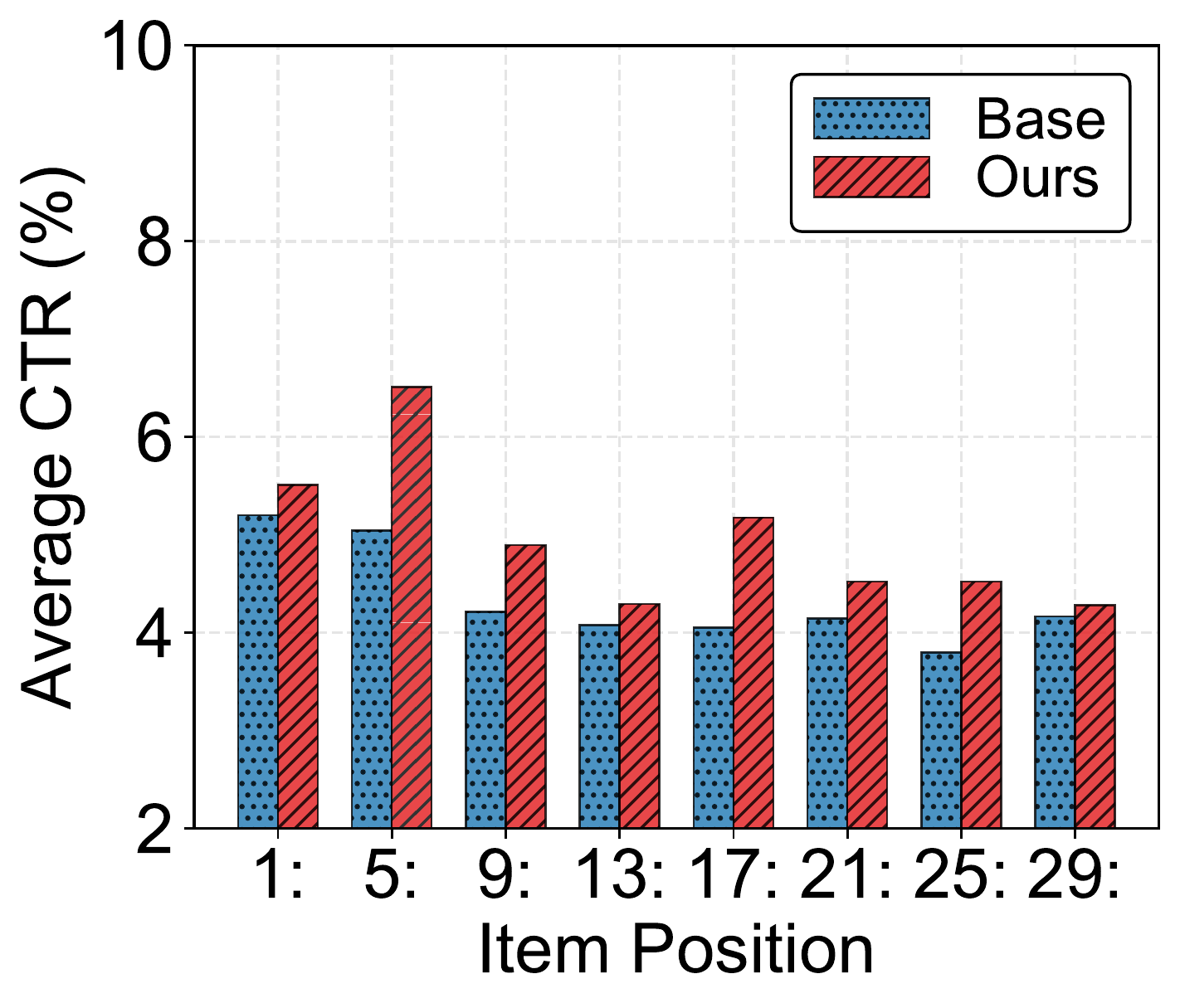}
		}
            \vskip -1em
		\caption{Online Performance Analysis.}
		\label{fig:onlinePerform}
	\end{figure}
	MIREC has been fully deployed in the homepage feed of Taobao named \textit{Guess-you-like} to serve the main traffic.
	Guess-you-like is one of the largest merchandise feed recommendation platform in China, which serves more than hundreds of millions of users toward billions of items every day.
	We deploy MIREC at the integrated recommendation stage in Guess-you-like platform, which takes hundreds of candidate items provided by multiple channels as input and outputs the final item list to return to the user.
	The online performance is compared against our previous baseline which is similar as a combination of $\beta$-WPO and HCA2E.
	In particular, the baseline uses a point-wise model for item quality estimation and uses a PID-based feedback control to automatically adjust parameter $ \beta $ to guarantee the  exposures for different channels. For each user request, the baseline also runs an MDP-based search method to determine the optimal card layout based on the estimated scores, which is similar as the heuristic search method in HCA2E.

	The overall performance in Table~\ref{tab:online} is averaged over two consecutive weeks.
	The results show that compared with the baseline method, MIREC brings an improvement of $ 3.00\% $ for CLICK, $ 1.75\%$ for CTR, $1.56\% $ for GMV, and $ 1.42\% $ for stay time. Compared with the fixed method, MIREC brings an improvement of $ 3.00\% $ for CLICK, $ 1.75\%$ for CTR, $1.56\% $ for GMV, and $ 1.42\% $ for stay time.
	These improvements indicate that our framework is able to increase user's willingness to stay and interact with the recommended items in practical applications. It is noteworthy that
	$ 1\% $ improvement on CLICK in Guess-you-like brings millions of clicks every day.
	Figure.~\ref{fig:onlinePerform} shows a detailed comparison of the exposure allocation results of a specific channel, where the items from this channel have a generally lower CTR than others.
	Each line in Figure.~\ref{fig:dailyPVR} represents the robustness of long-term exposure guarantee of this channel within two consecutive weeks.
	It is clear that compared with the baseline, our proposed MIREC is more robust to alleviate daily exposure fluctuations.
	The distribution of exposures on different positions in the feed is given in Figure.~\ref{fig:posPVR}.
	The result shows that our proposed framework tends to put lower-quality items backward to increase the overall utilities of all channels.
	Consequently, as shown in Figure.~\ref{fig:posCTR}, the averaged CTR of all channels on each position can be improved remarkably. This verifies that MIREC is superior in optimizing the item layout from a global perspective.

	\section{Conclusion}
        In this paper, we consider the integrated recommendation task with exposure constraints.
        In particular, we propose a two-layer framework named MIREC, where the allocation-layer optimizes the exposure of different channels, while the ranking-layer determines the optimal item layout on each page.
        Extensive experiments verified the effectiveness of our proposed framework.
        MIREC has been implemented on the homepage feed in Taobao to serve hundreds of millions of users towards billions of items every day.
 \bibliographystyle{plain}
\bibliography{MIREC_arxiv}

        \clearpage
        \newpage
        \appendix

        
        \section{Proof of Regret Bound}
        Our proof shares the same spirit as that of Theorem~1 in~\cite{balseiro2022best,lobos2021joint}. The difference is that \cite{balseiro2022best} does not consider a lower resource limit while~\cite{lobos2021joint} develops proof with an additional learnable parameter within $ f(\bx_t) $ and $ g(\bx_t) $. Therefore, we here develop a separate proof that is consistent with our formulation.
        We directly refer to a few propositions in~\cite{balseiro2022best,lobos2021joint} as preliminaries for simplicity. It is noteworthy that developing new proof of the online revenue maximization problem is not the main focus of this paper.

	Recall that the integrated recommendation problem is
	\begin{align}
		\mathcal{P}_{0}: \quad  &\text{OPT}(\mathcal{S}) = \max_{\bx_t \in \mathcal{X}} \sum\nolimits^T_{t=1} f(\bx_t) \\
		\mathrm{s.t.}  \quad & C_{1}: \sum\nolimits^T_{t=1} g_m(\bx_t) \leq G_{m,th}^{\max} N(\mathcal{S}), \forall m \in \mathcal{M}, \\
		& C_{2}: \sum\nolimits^T_{t=1} g_m(\bx_t) \geq G_{m,th}^{\min} N(\mathcal{S}), \forall m \in \mathcal{M}.
	\end{align}
	Since $ N(\mathcal{S}) $ denotes the sum exposures over the entire time horizon from $ t=1 $ to $ T $, we can replace the the upper exposure limit $ G_{m,th}^{\max} N(\mathcal{S}) $ and lower exposure limit $ G_{m,th}^{\min} N(\mathcal{S}) $ with $ TG_m $ and $ \alpha TG_m $ for simplicity, respectively, where $ G_m, \alpha \in [0, 1] $ are constants. 
	As such, problem~$ \mathcal{P}_0 $ can be reformulated as 
	\begin{align}
		\mathcal{P}_{1}: \quad  &\text{OPT}(\mathcal{S}) = \max_{\bx_t \in \mathcal{X}} \sum\nolimits^T_{t=1} f(\bx_t) \\
		\mathrm{s.t.}  \quad & \alpha T G_m \leq \sum\nolimits^T_{t=1} g_m(\bx_t) \leq T G_m, \forall m \in \mathcal{M}.
	\end{align}

	Before our analysis, we define constants $\bar{f}$, $\bar{g}>0$, $\underline{G}>0$ and $\bar{G}>0$ such that $\sup_{ \bx\in\mathcal{X}} f(\bx) \le \bar{f}$,  $\sup_{\bx\in\mathcal{X}} g(\bx) \le \bar{g}$, $\underline{G}:=\min_{m\in\mathcal{M}}G_m$ and $\bar{G}:=\max_{m\in\mathcal{M}}G_m$. Also, $\theta$ refers to the strongly-convexity parameter of the reference function $h(\cdot)$.
	
	First, we bound the dual iterates as follows.
	\begin{Assumption}[]
		\label{assumption}
		There exists a constant $ C_h > 0 $ such that the dual iterates $ \mu^t $ satisfy $ \mathbb{E}[||\nabla h(\mu^t)||_{\infty}] \leq C_h , \forall t \in [T] $.
	\end{Assumption}
	
	\begin{Remark} \label{rmk1}
		Note that, when choosing the reference function $h(\lambda):=\frac{1}{2}\|\lambda\|^2$, Assumption~\ref{assumption} can be omited according to Proposition~3 in~\cite{lobos2021joint}.
	\end{Remark}
	
	Denote the online Algorithm~\ref{algorithm} as $ \pi $ which makes a real-time decision $ \bx_t $ at time $ t $. Define the stopping time $ \tau_{\pi} \leq T $ as the minimum between $ T $ and the smallest time $ t $ such that there exists $ m \in \mathcal{M} $ with $ \sum_{t=1}^{\tau_{\pi}} g_m(\bx_t) + \bar{g} > T G_k  $. In other words, $ \tau_{\pi} $ refers to the first time the violation of one resource constraint happens. We bound the averaged gap between $ T $ and $ \tau_{\pi} $ as follows. 
	\begin{Proposition}\label{prop:Ttau}
		Suppose that Assumption~\ref{assumption} holds, using a constant step-size $ \eta > 0 $ in Algorithm~\ref{algorithm} yields
		\begin{align}
			\mathbb{E} \left[ T - \tau_{\pi} \right]  & \le \frac{\bar{g}}{\underline{G}}+\frac{C_h + \lVert \nabla h(\lambda^1) \rVert_\infty}{\eta \underline{G}}.
		\end{align}
	\end{Proposition}
	{\smallskip\noindent\it Proof.} 
	According to Step.~9 in Algorithm~\ref{algorithm} we have 
	\begin{equation}\label{key}
		\begin{aligned}
			\nabla{\mu}_{k,t}  &= - g_{k}(\bx_t) + G_{k} \left(\mathbbm{1}(\bmu_{k} \ge 0) + \alpha_{k} \mathbbm{1}(\bmu_{k} < 0)\right), \\
			&\leq -g_{k}(\bx_t) + G_{k},\quad \forall k \in [m].
		\end{aligned}
	\end{equation}
	Assume that the stopping time $ \tau_{\pi} $ is activated due to the violation of constraint on $ k $-th channel, we have
	\begin{align}
		\sum_{t=1}^{\tau_{\pi}} \nabla{\mu}_{k,t}  \leq & G_{k} \tau_{\pi} - \sum_{t=1}^{\tau_{\pi}}  g_{k}(\bx_t) 
		\le G_{k} \tau_{\pi} - T G_{k} +  \bar{g},
	\end{align}
	which leads to 
	\begin{align}\label{eq:T-tau}
		T - \tau_{\pi}  \le & \frac{1}{G_{k}}\left( \bar{g} - \sum_{t=1}^{\tau_{\pi}}  \nabla{\mu}_{k,t} \right).
	\end{align}
	According to Proposition~6 in~\cite{lobos2021joint}, the gradients of mirror descent satisfy
	$\nabla h_{k}(\mu_{k}^{t+1}) \ge \nabla h_{k}(\mu_{k}^t) - \eta \nabla{\mu}_{k,t}^t, \forall t \le \tau_{\pi}$, such that $ - \sum_{t=1}^{\tau_{\pi}}  \nabla{\mu}_{k,t} \le \frac{1}{\eta} \left( \nabla h_{k}(\mu_{k}^{\tau_{\pi}+1}) - \nabla h_{k}(\mu_{k}^1) \right) $. 
	Combing with the inequality in~\eqref{eq:T-tau}, we obtain
	\begin{align}
		\mathbb{E} \left[ T - \tau_{\pi} \right] & \leq \frac{\bar{g}}{G_{k}}+ \mathbb{E} \left[ \frac{\nabla h_{k}(\mu_{k}^{\tau_{\pi}+1}) - \nabla h_{k}(\mu_{k}^1)}{\eta G_{k}} \right]  \\
		& \leq \frac{\bar{g}}{\underline{G}}+\frac{C_h + \lVert \nabla h(\lambda^1) \rVert_\infty}{\eta \underline{G}},
	\end{align}
	as required. \hfill$\blacksquare$
	
	
	Let us denote the random variable $\gamma_t$ to be the type of the request at period $t$, which can determine the sample of the request. 
	\begin{Proposition}\label{Pro:CorrectRegretUpToTau_A}
		Using a constant step-size rule $\eta > 0$ for $ t>1 $ in Algorithm~\ref{algorithm}, it holds
		\begin{align}
			\mathbb{E}\!\left\lbrack  \tau_{\pi} D(\bar{\bmu}^{\tau_{\pi}}) \!-\! \sum_{t=1}^{\tau_{\pi}}  f(\bx_t)  \right\rbrack  \!\le\! & \frac{ 2 (\bar{g}^2 +\bar{G}^2)}{\theta} \eta \mathbb{E}[\tau_{\pi}] + \frac{1}{\eta} V_h(\bmu,\bmu^1),
		\end{align}
		where $\bar{\bmu}^{\tau_{\pi}} = \frac{\sum_{t=1}^{\tau_{\pi}} \bmu^t}{\tau_{\pi}}$.
	\end{Proposition}
	
	{\smallskip\noindent\it Proof.}
	According to the definition of $ \nabla \bmu_t $ and the subgradient inequality, we have
	\begin{align}
		&( \nabla \bmu^t )^T ( \bmu^t -  \bmu)  \ge   D( \bmu^t) - D( \bmu) \nonumber \\
		&\ge   D( \bmu^t) - \left( \mathbb{E}_{\gamma_t} [\varphi( \bmu)] + \sum_{k \in [K]} G_k([ \bmu_k]_+ - \alpha_k [- \bmu_k]_+)  \right),
	\end{align}
	where $ \varphi( \bmu) = f^*(\bmu) - \bmu^T g(\bx_t) $. 
	Considering that $\bx_t$ is an optimal solution of $\varphi(\bmu_t)$ not of $\varphi(\bmu)$, we have $ f(\bx_t) - \bmu^T g(\bx_t)  \le \varphi(\bmu) $. Then, by taking $\bmu = [0,0,\dots,0]$, and summing from one to $\tau_{\pi}$, we obtain
	\begin{align}
		& \sum_{t=1}^{\tau_{\pi}} ( \nabla \bmu^t )^T (\bmu^t - 0) \nonumber \ge \sum_{t=1}^{\tau_{\pi}}  D(\bmu^t) - \sum_{t=1}^{\tau_{\pi}} \mathbb{E}_{\gamma_t} [f(\bx_t)] \nonumber \\
		\ge & \tau_{\pi} D(\bar{\bmu}^{\tau_{\pi}}) - \sum_{t=1}^{\tau_{\pi}} \mathbb{E}_{\gamma_t} [f(\bx_t))],\label{Eq:AppProRegretUpToStable}
	\end{align}
	where $\bar{\bmu}^{\tau_{\pi}} = \frac{\sum_{t=1}^{\tau_{\pi}} \bmu^t}{\tau_{\pi}}$ and the inequality is based on the fact that the dual function is convex.
	In this paper, we adopt $ \theta $-strongly convex function as the relation function $ h(\cdot) $ in mirror descents.
	According to Step.~2 of Proposition~8 in~\cite{lobos2021joint}, we have
	\begin{equation}\label{eq:prelimBound}
		\mathbb{E}\left[\sum_{t=1}^{\tau_{\pi}} (\nabla \bmu^t)^T(\bmu^t-\bmu) \right] \leq \frac{ 2 (\bar{g}^2 +\bar{G}^2)}{\theta} \eta \mathbb{E}[\tau_{A}] + \frac{V_h(\lambda,\lambda^1)}{\eta}.
	\end{equation}
	Combining~\eqref{eq:prelimBound} with~\eqref{Eq:AppProRegretUpToStable}, we get 
	\begin{equation}\label{key}
		\mathbb{E}\!\left\lbrack  \tau_{\pi} D(\bar{\bmu}^{\tau_{\pi}}) \!-\! \sum_{t=1}^{\tau_{\pi}}  \mathbb{E}_{\gamma_t} [f(\bx_t)]  \right\rbrack  \!\le\! \frac{ 2 (\bar{g}^2 +\bar{G}^2)}{\theta} \eta \mathbb{E}[\tau_{\pi}] + \frac{1}{\eta} V_h(\bmu,\bmu^1).
	\end{equation}
	According to Step.~3 of Proposition~8 in~\cite{lobos2021joint},
	we have 
	\begin{equation}\label{eq:martingle}
		\mathbb{E}\!\left\lbrack \sum_{t=1}^{\tau_{\pi}}  \mathbb{E}_{\gamma_t} [f(\bx_t)]  \right\rbrack = \mathbb{E}\!\left\lbrack \sum_{t=1}^{\tau_{\pi}}  f(\bx_t)  \right\rbrack.
	\end{equation}
	Combining~\eqref{key} and \eqref{eq:martingle}, we complete the proof of Proposition \ref{Pro:CorrectRegretUpToTau_A}. ~\hfill$\blacksquare$

	Before providing more details on the proof of regret bound, we introduce a new benchmark of problem $ \mathcal{P}_0 $ as in~\cite{lobos2021joint} due to that problem $ \mathcal{P}_0 $ may be infeasible due the presence of both lower and upper exposure constraints. Specifically, we define
	\begin{align*}
		\mathrm{F}(\bx_t,\lambda) &= (1-\lambda)f(\bx_t)  + \lambda\mathbb{E}_{\cS}[f(\bx_t)] \\
		\mathrm{G}(\bx_t,\lambda) &= (1-\lambda)g(\bx_t) + \lambda \mathbb{E}_{\cS}[g(\bx_t)],
	\end{align*}
	where $\lambda \in [0,1]$ is the interpolation parameter. We define
	\begin{align}
		\mathrm{OPT}(\cS,\lambda) = \nonumber 
		\mathbb{E}_{\cS^T}
		\left\lbrack
		\begin{array}{c l}	
			\underset{\bx^t, t \in [T]}{\max} \ \ \sum\nolimits_{t=1}^T F(\bx_t,\lambda)  &  \\
			\textrm{s.t.} \ \   T \alpha \odot G \le \sum_{t=1}^T G(\bx_t,\lambda) \le TG 
		\end{array}
		\right\rbrack \nonumber 
	\end{align}
	where $\cS^T := \cS \times \dots \times \cS$ is a product distribution of length $T$.
	Now we give the definition of the new benchmark as 
	\begin{equation}
		\mathrm{OPT}(\cS) := \max_{\lambda \in [0,1]}  \mathrm{OPT}(\cS,\lambda). \label{Eq:OptOffline}
	\end{equation}
	This benchmark is an interpolate value between the expected optimal value of problem $ \mathcal{P}_0 $ and a deterministic problem which replaces the varying utility values $ f(\bx_t) $ and cost values $ g(\bx_t) $ with their expected values. 
	For this benchmark, we have
	\begin{equation}
		\begin{aligned}\label{eq:bound-opt}
			\EE_{\cS} \left[ \OPT(\cS) \right] &=  \frac{\tA}{T} \EE_{\cS} \left[ \OPT(\cS) \right] + \frac{T-\tA}{T}\EE_{\cS} \left[ \OPT(\cS) \right] \\
			& \le  \tA \bar D(\bmu_{\tA}| \cS) + \pran{T-\tA}{\ubf},
		\end{aligned}
	\end{equation}
	where the inequality uses the fact that $ \OPT(\cS) \leq D(\bmu|\cS)$ according to Proposition~1 in~\cite{lobos2021joint}and that $\OPT(\cS)\le T \ubf$. 
	Combining all findings together, we have
	\begin{subequations}
		\begin{align}
			&\Regret{\pi|\cS} = \EE_{\cS}\left[ \OPT(\cS)-R(\pi|\cS) \right] \\
			& \le \EE_{\cS}\left[\tA \bar D(\bmu_{\tA}| \cS) + \pran{T-\tA}{\ubf} - \sum_{t=1}^{\tA} f(\bx_t) \right] \\
			& = \EE_{\cS} \left[ \tA \bar D(\bmu_{\tA}| \cS) - \sum_{t=1}^{\tA} f(\bx_t)\right] +  \EE_{\cS} \left[ T-\tA \right]\ubf \\  
			& \le \frac{ 2 (\bar{g}^2 +\bar{G}^2)}{\theta} \eta \mathbb{E}[\tau_{\pi}] + \frac{1}{\eta} V_h(\bmu,\bmu^1) + \frac{\bar{g}}{\underline{G}}+\frac{C_h + \lVert \nabla h(\lambda^1) \rVert_\infty}{\eta \underline{G}},
		\end{align}%
	\end{subequations}
	where the first inequality is from \eqref{eq:bound-opt} and the second inequality is from Proposition~\ref{prop:Ttau} and Proposition~\ref{Pro:CorrectRegretUpToTau_A}.
	Therefore, the constants in Theorem~1 are $ C_1 = \frac{\bar{g}}{\underline{G}} $, $ C_2 = \frac{ 2 (\bar{C}^2 +\bar{b}^2)}{\theta} \eta $, and $ C_3 = V_h(\bmu,\bmu^1) + \frac{\bar{g}}{\underline{G}}+\frac{C_h + \lVert \nabla h(\lambda^1) \rVert_\infty}{\underline{G}} $, respectively. Moreover, recall Remark~\ref{rmk1}, we choose $h(\lambda):=\frac{1}{2}\|\lambda\|^2$ and dual iterates are bounded. Hence, we complete the proof of Theorem~\ref{thm1}. ~\hfill$\blacksquare$

	\section{Proof of Cost Feasibility}
	Proposition~\ref{prop:lower_bound} shows that a solution obtained using Algorithm~\ref{algorithm} can not overspend, but may underspend. 
	Based on the definition of subgradient $\nabla \mu_{k}^{t}$, we have
	\begin{equation}
		\begin{aligned}
			\frac{\nabla h_k(\mu^1) \!-\! \nabla h_k(\mu^{\tau_{\pi}+1})}{\eta} \!=\! \sum_{t=1}^{\tau_A} (G_k(\mathbbm{1}(\mu_k \!\ge\! 0) \!+\! \alpha_k \mathbbm{1}(\mu_k \!<\! 0)) \!-\! g_k(\bx_t)) \label{eq:cost1}
		\end{aligned}
	\end{equation}
	Now, given that $\mathbbm{1}(\mu \ge 0) +\alpha_k \mathbbm{1}(\mu <0) \ge \alpha_k$ for any $\mu \in \bbR$ and that $\tau_{\pi} \le T$ by definition, we have
	\begin{equation}
		\begin{aligned}
			\sum_{t=1}^{\tau_A} G_k\left(\mathbbm{1}(\mu_k \!\ge\! 0) \!+\! \alpha_k \mathbbm{1}(\mu_k \!<\! 0)\right) + (T \!-\! \tau_A ) \alpha_k b_k  \!\ge\! T\alpha_k b_k.\label{eq:cost2}
		\end{aligned}
	\end{equation}
	Combining~\eqref{eq:cost1} and~\eqref{eq:cost2} and taking expectation, we get
	\begin{subequations}
		\begin{align}
			&T \alpha_k G_k - \mathbb{E}[\sum_{t=1}^{\tau_{\pi}}g_k(\bx_t)]  \\
			&\le \frac{\nabla h_k(\mu^1) - \mathbb{E}[\nabla 	h_k(\mu^{\tau_{\pi}+1})]}{\eta} + \mathbb{E}[T-\tau_A] \alpha_k b_k \\
			& \le   \left( \frac{\lVert \nabla h(\lambda^1) \rVert_\infty + C_h}{\eta } \right) \frac{\underline{G} + \alpha_k b_k}{\underline{G}} +  \frac{\alpha_k b_k \bar{g}}{ \underline{G}},
		\end{align}
	\end{subequations}
	where the second inequality comes from Proposition~\ref{prop:Ttau}.
	
	%

\end{document}